\documentclass[12pt,letterpaper]{article}                  
\usepackage{achemso}
\usepackage{overcite,hyperref}
\usepackage{graphicx}
\usepackage{soul}
\usepackage{color}

\def\biexciton{BX~}
\def\exciton{X}
\definecolor{red}{rgb}{0,0,1}
\begin{document}

\title{A Large Blue Shift of the Biexciton State in Tellurium Doped CdSe Colloidal Quantum Dots}

\author{Assaf Avidan and Dan Oron}

\maketitle

\begin{abstract}
The exciton-exciton interaction energy of Tellurium doped CdSe
colloidal quantum dots is experimentally investigated. The dots
exhibit a strong Coulomb repulsion between the two excitons, which
results in a huge measured biexciton blue shift of up to $300$~meV.
Such a strong Coulomb repulsion implies a very narrow hole wave
function localized around the defect, which is manifested by a large
Stokes shift. Moreover, we show that the biexciton blue shift
increases linearly with the Stokes shift. This result is highly
relevant for the use of colloidal QDs as optical gain media, where a
large biexciton blue shift is required to obtain gain in the single
exciton regime.
\end{abstract}

\section*{}

The progress in chemical synthesis of colloidal quantum dots,
provides a way to produce high quantum yield nanocrystals out of a
variety of substances, and allows control over their size, shape and
composition. Because of the strong charge confinement within the
dots, their electronic excitation spectrum is discrete, with an
increased energy gap as compared to its corresponding bulk material.
In the strong confinement regime, the band gap is conveniently tuned
simply by changing the QD diameter. This particular property of
colloidal small QDs serves as the prime motivation to further
investigate and optimize their properties as optical gain media. The
first successful attempt to measure optical gain from colloidal
QDs\cite{bawendi_science} showed that for CdSe, optical
amplification requires high power density pumping. This is due to
the degeneracy between the exciton (\exciton)~and biexciton (BX)
resulting from a small \exciton-\exciton~interaction in a doubly
excited QDs. Since the electronic ground state is doubly degenerate,
the threshold for population inversion requires that at least some
of the dots are populated with two electron-hole pairs. However, the
\biexciton state decays through fast non-radiative Auger
recombination, during which one electron-hole pair recombines and
transfers its energy to a spectator charge carrier. Thus, in the
strong confinement regime, the gain dynamics is dominated by a fast
non-radiative decay, which is typically of the order of tens of
picoseconds. This system is therefore impractical for achieving
sustained gain from QD-based devices. In order to obtain optical
gain from singly excited colloidal QDs, it is thus necessary to
remove the \exciton-\biexciton degeneracy. In core only QDs, because
of the Coulomb correlation term between the two excitons, the
\biexciton state is bound\cite{koch}. This implies that the
\biexciton is red shifted in energy with respect to the energy of
two noninteracting excitons. This red shift is however, typically
$\leq$ $30$~meV, which is less than the inhomogeneous line
broadening of the \exciton~and \biexciton states. Therefore the
degeneracy is effectively not removed.

Recently it was demonstrated that in core/shell type II
nanocrystals, the core and shell sizes can be changed in order to
tune the relative position of the \exciton~and \biexciton
lines\cite{oron_typeii,klimov_typeii}. These dots are composed of
two different materials with a staggered band diagram
\cite{bawendi_type_two}, such that the holes tend to localize in one
material and the electrons tend to localize in the other. This
charge separation leads to a strong Coulomb repulsion that dominates
the \exciton-\exciton~interaction, resulting in a blue shift of the
\biexciton emission line \cite{oron_typeii}. Such a system was
realized experimentally, by using CdS/ZnSe core/shell QDs
\cite{klimov_typeii}, and a large \biexciton blue shift of $100mev$
was measured. This value is large enough to overcome the
inhomogeneous broadening of the QDs ensemble. Therefore, the
degeneracy is almost completely removed, resulting in a
significantly reduced gain threshold. It should be noted, however,
that in ref.~\cite{klimov_typeii}, the core and shell parameters and
composition were chosen to maximize the \biexciton blue shift
\cite{klimov_typeii_analysis} while still having a large enough
transition moment. Thus, such a system lacks the robustness required
for broad band color tunability.

\begin{figure}
\centering
\includegraphics[width=12cm]{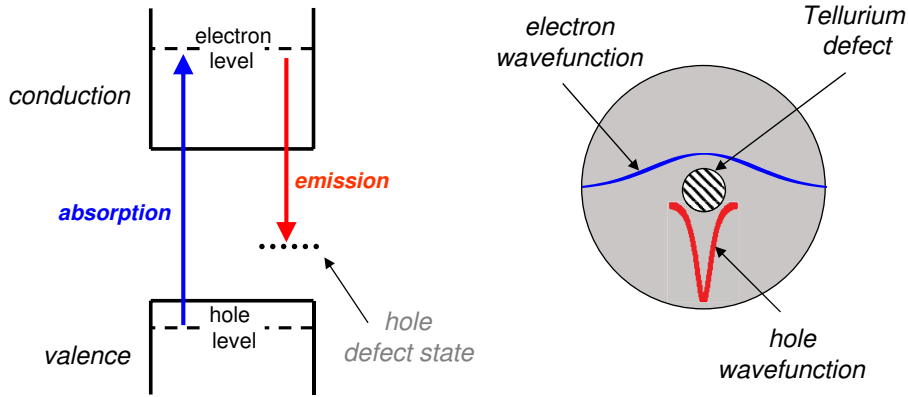}
\caption{Schematics of a doped QD. Left: The energy band structure
is composed of the energy structure of the host nanocrystal, plus an
additional level which resides inside the host band gap. Right:
spatial structure of doped QDs. The electron wave function (blue)
width is determined by the size of the host nanocrystal. The hole
wave function (red) is spatially localized around the Te defect.}
\label{doped_draw}
\end{figure}

 Here we suggest an alternative way to control the
Coulomb repulsion between two \exciton s inside a QD, by
incorporating dopant atoms into a host nanocrystal. This method has
already shown to induce large \exciton-\exciton~repulsion in an
organic-inorganic two dimensional semiconductor
system\cite{bulk_doping}. Although widely used in bulk
semiconductors, doping of colloidal nanocrystals became a practical
possibility only recently. Doped nanocrystals exhibit a large Stokes
shift between the absorption edge and their emission, since
absorption is determined by the (large) host band gap, while
emission is due to defect states inside the gap (see schematic
representation in Figure~\ref{doped_draw}). Some successful examples
include Mn doped ZnSe \cite{peng2,norris_first}, $Cu$ doped ZnSe
\cite{peng1}, and Mn doped CdS \cite{white_light} (see also a review
by Norris\cite{norris_review}). Here we investigate the effect of Te
doping on the optical properties of the \exciton~and \biexciton
states in colloidal CdSe nanocrystals. In analogy to type II
CdTe/CdSe QDs where the \biexciton repulsion is due to charge
separation \cite{oron_typeii}, we have chosen to work with doped QDs
in which the dopant atoms form a defect state, which spatially
localizes only the holes~\cite{weller}. We show that a few atoms of
Te incorporated in each dot, can have a significant effect not only
on the linear \cite{weller} but also on the nonlinear properties of
the entire nano-composite.

The synthesis of Te doped CdSe (CdSe:Te) was performed by following
the procedure suggested by Franzl et al.\cite{weller} with minor
modifications. The QDs were synthesized in a non coordinating
solvent (octadecene) with TDPA as a ligand replacing TOPO and HDA as
in ref. \cite{weller}. The Se and Te precursors were dissolved in a
single 2 ml TOP solution, with Te percentage ranging from 0-12\%,
and injected into a vigorously stirred $Cd$ solution at a
temperature of $300\,^{\circ}\mathrm{C}$. The particles were grown
at a temperature of $250^o C$, and aliquots were taken at several
time intervals corresponding to particle diameters in the range
$2-4$~nm. Further details can be found in the supporting online
information.

\begin{figure}
\centering
\includegraphics[width=12cm]{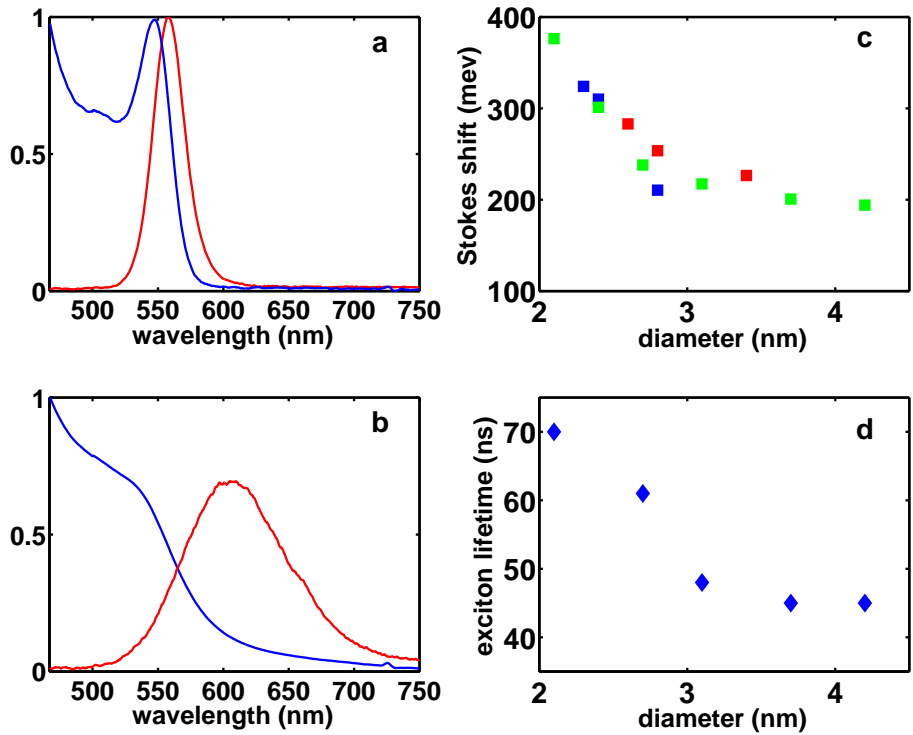}
\caption{(a-b) Absorption (blue) and emission (red) spectra of
undoped CdSe (a), and CdSe:Te~$5\%$ (b) QDs, with similar average
diameter. For the doped dots, the absorption peak is smeared because
of size inhomogeneity and the band edge emission is substantially
broader. c) Stokes shift vs. dot diameter for CdSe:Te$3\%$ (blue),
CdSe:Te$5\%$ (green) and CdSe:Te$7\%$ (red) QDs. d)
\exciton~lifetime vs. dot diameter for the CdSe:Te$5\%$
QDs.}\label{linear}
\end{figure}

Typical absorption and emission spectra of a colloidal solution of
CdSe:Te{$5\%$} ($5\%$ Te content in the injected solution) is shown
in Figure~\ref{linear}b, along with those of undoped CdSe QDs
(Figure~\ref{linear}a) prepared under similar conditions. The
emission spectrum was taken using a $405$~nm CW laser, and the
absorption measurements were carried with a UV-vis
spectrophotometer. The absorption curve shows the first electronic
excitation of the QDs at about $540$~nm, where for the CdSe:Te{5\%}
QDs (with the same mean size as the undoped CdSe) this peak is
partially smeared because of a broader size distribution. The
emission of the undoped QDs reveals the familiar emission profile of
CdSe featuring a small Stokes shift, while the doped CdSe exhibits a
large Stokes shift of about $60$~nm and a much broader emission
distribution. The emission profile shown in Figure~\ref{linear}b for
the CdSe:Te{5\%}, is in complete agreement with previous results
\cite{weller}.

In ref.~\cite{weller} it was shown that for QDs with a fixed size,
the emission wavelength is further red shifted as a function of the
amount of Te injected during the synthesis, while the absorption
onset remains unchanged. As discussed in ref.~\cite{weller}, the red
shifted emission results from a hole trap state inside the CdSe gap,
which is caused by few atoms of Te bunched together during the
nucleation. The absorption onset is determined by the energy gap of
the host material, and its position is hardly affected by the
dopant. The size distribution is, however, clearly affected by the
introduction of the Te atoms, especially at the early stages of the
growth.

In addition to the dependence of the Stokes shift on the amount of
Te incorporated into the CdSe Qds, as presented in
ref.~\cite{weller}, it is interesting to further investigate its
dependence on the QDs size. Since the energy level of the hole trap
state is expected to weakly depend on size, the Stokes shift should
follow the spectral position of the hole ground state in the host
nanocrystal. Figure~\ref{linear}c shows the CdSe:Te{5\%} QD Stokes
shift as a function of the dot diameter. For the two largest dots,
the absorption peak is smeared, so we define it as the point where
the slope of the absorption decreases, after the onset of the band
edge absorption. As expected, the Stokes shift decreases as a
function of the crystal size. For the $2.1$~nm diameter dots (the
smallest dots from which band edge, rather than trap state,
fluorescence could be measured) the Stokes shift is as high as
$380$~meV. For large dots ($4$~nm diameter) it approaches a value of
about $180$~meV which corresponds the energy difference between the
Te defect level and the bulk CdSe valence band edge. On the same
graph, we plot the Stokes shift which results from the synthesis of
QDs with other contents of Te, namely CdSe:Te{$7\%$} (red)
CdSe:Te{$3\%$} (blue). As can be seen, a larger Te percentage upon
injection, results in a somewhat larger Stokes shift. However, we
have found that working above $7\%$ of Te is difficult due to the
nucleation of pure $CdTe$ QDs.

In order to observe emission dynamics we excited a dilute QD
solution in toluene with $4$~ns pulses at $450$~nm, at a repetition
rate of $10$ Hz. The emission was spectrally filtered by a
monochromator and transient emission traces were recorded with a
fast detector and a digital oscilloscope. When the excitation level
is low enough \cite{oron_typei} ($I_{input}<<1$photon/pulse/dot) we
observe only the \exciton~(linear) dynamics. Multiexciton transient
spectra were measured at higher photon fluences ($I_{input}>>1$
photon/pulse/dot). A more detailed description of the set up and the
detection scheme can be found elsewhere
\cite{oron_typei,oron_typeii}. In Figure~\ref{linear}d we present
measurements of the \exciton~ lifetime of a colloidal CdSe:Te{5\%}
solution, as a function of the QDs diameter. The results clearly
show that for small dots the lifetime decreases as a function of
size, while for larger dots it only weakly depends on size,
similarly to undoped CdSe QDs. Surprisingly, this result is opposite
to the lifetime size dependence measured on both type II CdTe/CdS
QDs \cite{oron_typeii}, and on quasi type II CdSe/CdS QDs, where the
hole wave function is localized and the electron is extended
\cite{quasi_type_ii}.
In type II QDs the lifetime increases continuously from about
$30$~ns for core only QDs, to about $150$~ns for a $2.5$~nm thick
shell. In such a core/shell system, the electron wave function is
spread over the entire shell, while the hole wave function is almost
unaffected by the shell growth, leading to a smaller electron-hole
overlap for thicker shells. In the quasi type two case the wave
functions overlap is less sensitive to the shell thickness, and the
exciton lifetime increases only slightly with the shell size.
\cite{quasi_type_ii}
 The lifetime measurements presented in
Figure~\ref{linear} imply that this picture is not correct in our
case. Here, the depth of the Te hole trap is determined by the
energy difference between the trap and the host valence band edge,
which is by definition the \textit{Stokes shift}. Thus, as the CdSe
particle grows, the hole becomes less localized in the trap. This
results in a larger overlap with the electron, despite the fact that
the electron is spread over a larger volume. As we show below, this
picture is consistent also with the measured \biexciton size
dependent spectra.

\begin{figure}
\centering
\includegraphics[width=12cm]{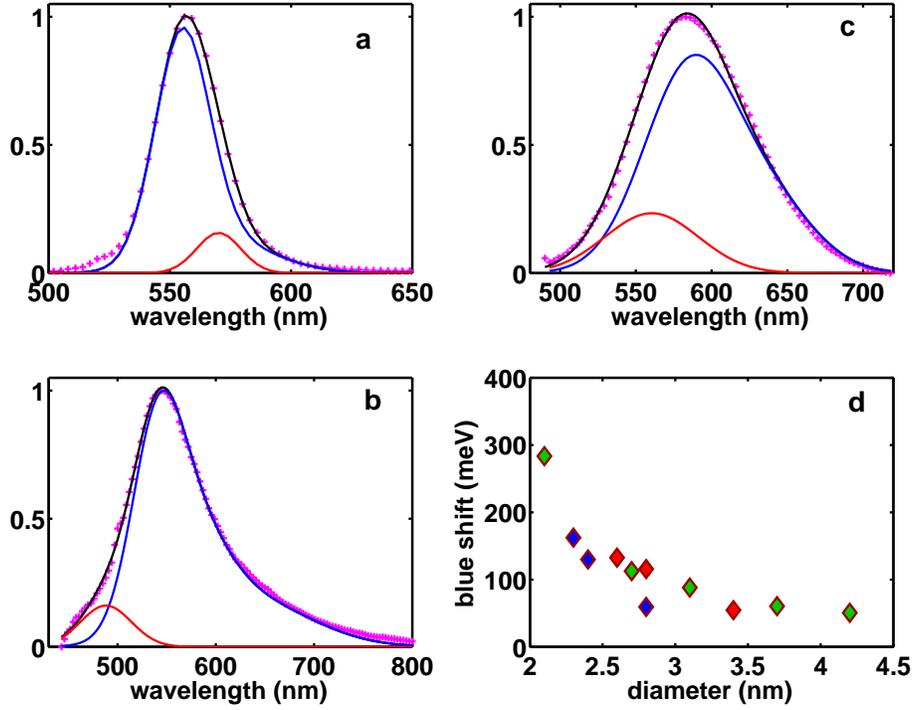}
\caption{ \biexciton spectroscopy of CdSe:Te{5\%} QDs.
 (a-c) Experimental transient emission spectra are shown in
magenta pluses along with its numerical fit (black). We decompose
the fit into a low power \exciton~profile (blue) and an additional
Gaussian peak to account for the \biexciton emission (red). a)
undoped $3$~nm diameter CdSe. b) $2.1$~nm diameter CdSe:Te{5\%}. c)
$2.7$~nm diameter CdSe:Te{5\%}. d) \biexciton~blue shift vs. dot
diameter for CdSe:Te{3\%} (blue), CdSe:Te{5\%} (green) and
CdSe:Te{7\%} (red) doped QDs. Adding more than $5\%$ Te does not
increase the blue shift substantially. }\label{nonlinear}
\end{figure}

We now turn to the measurements of the \exciton~and \biexciton
spectra in CdSe:Te dots. The transient spectra measurements are
taken at the peak of the excitation pulse, assuming that the Auger
lifetime is much shorter than the pulse duration (as we verify with
transient photoluminescence measurements, not shown). Spectra
measured at low photon fluence provide the line shape of the
\exciton, to which we add a Gaussian peak to numerically fit the
width and position of the \biexciton peak in the high power regime.
In Figure~\ref{nonlinear}a we plot the measured transient spectrum
of undoped $3$~nm diameter CdSe, along with its fit as described
above. On top we plot the break up of the fitted spectrum into its
\exciton~and \biexciton components. In Figure~\ref{nonlinear}$b$ and
Figure~\ref{nonlinear}$c$ we plot similar spectra of CdSe:Te{5\%}
QDs with diameters $d=2.1$~nm and $d=2.7$~nm respectively. The dots
in Figure~\ref{nonlinear}a and Figure~\ref{nonlinear}c have
approximately the same absorption onset (at about $530$~nm), while
the dots in Figure~\ref{nonlinear}a and Figure~\ref{nonlinear}b,
have the same emission peak wavelength of about $540$~nm. In
agreement with previous reports \cite{wogon}, the \biexciton line of
the undoped CdSe QDs is found to be about $30$ meV red shifted
compared to the \exciton~peak. In contrast, we have found that for a
CdSe:Te{5\%} QDs, the \biexciton is substantially blue shifted. In
fact, for the smallest dot (see Figure~\ref{nonlinear}b), at the
peak wavelength of the \exciton~emission, the \biexciton content is
vanishingly small. In Figure~\ref{nonlinear}d we plot the \biexciton
blue shift as a function of the dot diameter. As can be seen the
blue shift is larger for small QDs, approaching a remarkably large
value of $300$ meV, where for larger dots it saturates at about $50$
meV. The BX blue shift values of the CdSe:Te{$7\%$} and
CdSe:Te{$3\%$} are plotted as well on Fig.~\ref{nonlinear}d, where
it is seen that working above $5\%$ Te content does not
significantly improve the results.

This extraordinary blue shift is of the same order as the Te defect
state offset (Stokes shift), and is explained by strong charge
confinement around the Te atoms. The blue shift of the smaller dots
is even larger than that of an optimized type II system
\cite{klimov_typeii}, and therefore indicates a much stronger
confinement of the holes. This leads to a huge unscreened local
electric field which Stark shifts the \biexciton energy level.

\begin{figure}
\centering
\includegraphics[width=10cm]{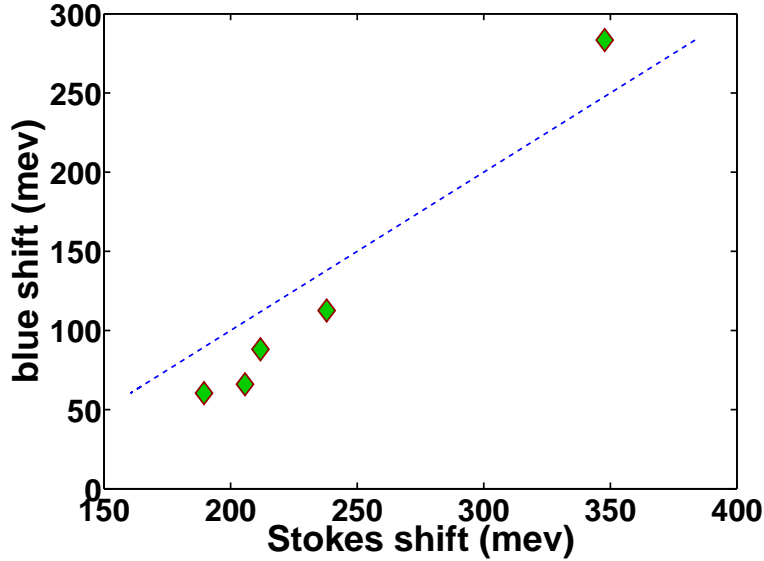}
\caption{\biexciton~blue shift as a function of the Stokes shift for
CdSe:Te$5\%$ doped QDs. Green diamonds: experimental data; dashed
blue line: Stokes shift minus $100$~meV as a function of the Stokes
shift itself. The blue shift roughly follows the size dependence of
the Stokes shift, with an additional offset of about
$-100$~meV.}\label{blue-stokes}
\end{figure}

The size dependence of the \biexciton blue shift shown in
Figure~\ref{nonlinear}d, resembles that of the Stokes shift shown in
Figure~\ref{linear}c. Following the arguments given to explain the
size dependence of the \exciton~lifetime, a large Stokes shift
implies a narrower hole wave function. Since the Coulomb repulsion
between the holes is determined by the local positive charge
density, we expect the blue shift to be an increasing function of
the Stokes shift. In Figure~\ref{blue-stokes} we plot the \biexciton
blue shift versus the corresponding Stokes shift. The dashed blue
line represents the Stokes shift minus $100$~meV as a function of
the Stokes shift itself, and is plotted as a guide. The dependence
of the blue shift on the Stokes shift is roughly linear with a slope
of about unity, implying that the \biexciton repulsion is pinned to
the Stokes shift. The agreement between the guide and the
measurements suggests that independently of size, the \biexciton
level always lies approximately $100$~meV below the host valence
band edge. In other words, the holes in the \biexciton state are
weakly localized to the same extent, independent of the dot
diameter. It is important to note that unlike type II QDs
\cite{klimov_typeii_analysis}, for CdSe:Te QDs the \biexciton blue
shift cannot be calculated simply by using perturbation theory.
Since the \biexciton Coulomb repulsion we measure is of the same
order as that of the Stokes shift (see Figure~\ref{blue-stokes}),
the wave functions of the holes in the \exciton~and \biexciton
states differ significantly. This has to be treated self
consistently in order to theoretically account for the large value
of the \biexciton blue shift. Thus, we believe that an exact
description of such a system in which one charge carrier is
spatially confined to a defect state inside the gap, poses a highly
relevant theoretical question.

Although the \biexciton blue shifts we have measured in CdSe:Te QDs
are huge, these dots exhibit a broader emission line than undoped
CdSe QDs prepared by the same method (compare Figure~\ref{linear}a
and Figure~\ref{linear}b). As is evident from the broadening of the
absorption peaks (see Figure~\ref{linear}b) this inhomogeneous
broadening is partially the result of a broad size distribution of
the CdSe host QDs. In addition, fluctuations in the number of Te
atoms in each dot and their specific location inside the dot may
contribute to this inhomogeneity as well. Similar to the results
presented in ref.~\cite{klimov_typeii} for type II CdS/ZnSe, the
results presented here show only a partial removal of the
\exciton-\biexciton degeneracy. However, this degeneracy can be more
efficiently removed either by an improved synthetic procedure or by
choosing different materials and morphology.

 To conclude, we have performed a multiexciton analysis of Te doped CdSe QDs. Unlike core
only QDs, we showed that the \biexciton is blue shifted with respect
to the \exciton~emission line. Moreover, We found that the
\biexciton blue shift is linearly proportional to the Stokes shift
with a slope of approximately unity. Our results imply that pushing
the defect state deeper inside the gap, will further increase the
coulomb repulsion between the holes in a \biexciton state. Another
surprising property of Te doped CdSe QDs is that their \biexciton
blue shift as well as their \exciton~lifetime reveal an opposite
size dependence than that of an analogous type II system. In doped
QDs the small size of the defect can lead to an extremely large
Coulomb repulsion of about $300$~meV, which is of the order of the
band offsets. This regime is dramatically different than that
observed in core/shell particles, opening new possibilities of
controlling \exciton-\exciton~in quantum confined systems.


\begin{thebibliography}{}


\bibitem{bawendi_science} Klimov, V. I.; Mikhailovsky, A. A.; Xu, S.; Malko, A.; Hollingsworth, J. A.; Leatherdale, C.A.; Eisler, H.-J.; Bawendi, M. G. \textit{Science} \textbf{2000}, \textit{290}, 314. 

\bibitem{koch} Banyai, L.; Koch, S. W. \textit{Semiconductor Quantum Dots} World Scientific, 1993.

\bibitem{oron_typeii} Oron, D.; Kazes, M.; Banin, U. \textit{Phys. Rev. B} \textbf{2007}, \textit{75}, 035330.

\bibitem{klimov_typeii} Klimov, V. I.; Ivanov, S. A; Nanda, J.; Achermann, M.; Bezel, I.; McGuire, J. A.; Piryatinski, A. \textit{Nature} \textbf{2007},  \textit{447}, 441. 

\bibitem{bawendi_type_two} Sungjee, K.; Fisher B.; H.-J. Eisler; Bawendi, M. \textit{J. Am. Chem. Soc.} \textbf{2003}, \textit{125}, 38, 11467. 

\bibitem{klimov_typeii_analysis} Piryatinski, A.; Ivanov, S. A.; Tretiak, S.; Klimov, V. I. \textit{Nano Lett.} \textbf{2007}, \textit{7}, 108. 

\bibitem{bulk_doping} J.-I. Fujisawa; T. Ishihara \textit{Phys. Rev. B} \textbf{2004}, \textit{70}, 205330. 

\bibitem{peng2} Pradhan, N.; Peng, X. \textit{J. Am. Chem. Soc.} \textbf{2007}, \textit{129}, 11.

\bibitem{norris_first} Norris, D. J.; Yao, N.; Charnock, F. T.; Kennedy, T. A. \textit{Nano Lett.} \textbf{2001}, \textit{1}, 1. 

\bibitem{peng1} Pradhan, N.; Goorskey, D.; Thessing, J.; Peng, X \textit{J. Am. Chem. Soc.} \textbf{2005}, \textit{127}, 17586.


\bibitem{white_light} Nag, A.; Sarma, D. D. \textit{J. Phys. Chem. C} \textbf{2007}, \textit{111}, 37.

\bibitem{norris_review} Norris, D. J.; Efros, A. L.; Erwin, S. C. \textit{Science} \textbf{2008}, \textit{319}, 28.

\bibitem{weller} Franzl, T.; M\"uller, J.; Klar, T. A.; Rogach, A. L.; Feldmann, J.; Talapin, D. V.; Weller, H. \textit{J. Phys. Chem. C} \textbf{2007}, \textit{111}, 2974. 

\bibitem{oron_typei} Oron, D.; Kazes M.; Banin, U. \textit{Phys. Rev. B} \textbf{2006}, \textit{74}, 115333. 

\bibitem{quasi_type_ii} Zeng, Q.; Xianggui, Q.; Yajuan, S.; Youlin, Z.; Langping, T.; Jialong, Z.; Hong, Z.  \textit{J. Phys. Chem. C} \textbf{2008}, \textit{112}, 8587–8593. 

\bibitem{wogon} Woggon, U. in \textit{Optical Properties of Semiconductors and Their Nanostructures}, edited by H. Kalt and M. Hetterich, Springer-Verlag: Berlin, \textbf{2004}.




%
%
%
%
%
%
%
%
%
%
%









\end{thebibliography}
\end{document}